\documentstyle[11pt,newpasp,twoside,epsf]{article}
\def\sm{M$_\odot$}
\def\zmedio{$\langle$$|z|$$\rangle$}
\def\bpn{$b$ PNe}
\def\epn{$e$ PNe}

\def\edcomment#1{\iffalse\marginpar{\raggedright\sl#1\/}\else\relax\fi}
\reversemarginpar

\begin{document}
\title{Morphology vs. physical properties: some comments and questions}
 \author{Romano L.M. Corradi}
\affil{Instituto de Astrof\'\i sica de Canarias, E-38200 La Laguna, Tenerife, 
Spain}

\begin{abstract}
Some of the correlations between morphological and other physical
properties of planetary nebulae (PNe) are reviewed.

In particular, the finding that bipolar ($b$) PNe have more massive
progenitors than the other morphological classes is discussed in detail.
Earlier results are confirmed; including all the various sources of
uncertainty, the Galactic distribution of objects indicate that \bpn\ are
formed by stars with initial masses $>1.3$~\sm\ while elliptical ($e$) PNe by
progenitors with masses $<1.3$~\sm.

Recent results for the chemical abundances of $b$ and $e$ PNe and their
orientation within the Galaxy are also presented.

Finally, the key role in the discussion of the formation of \bpn\ played by
detached binary systems such as symbiotic stars is briefly discussed.
\end{abstract}

\section{Introduction}

It is now clear that a classification of PNe based on their morphology
(Balick 1987; Schwarz, Corradi, \& Stanghellini 1993; Manchado et al.
 1996) also corresponds to a real physical classification of the
nebulae and of their stellar progenitors.  Correlations between morphology
and other properties of the nebulae/stars were recognized quite a long time
ago (cf. Greig 1972; Peimbert \& Torres-Peimbert 1983; Zuckerman \& Gatley
1988), and confirmed recently by the analysis of extensive (and thus
statistically more robust), homogeneous, and high-quality image atlases
(Balick 1987; Schwarz, Corradi, \& Melnick 1992; Manchado et al. 1996; Gorny
et al. 1999).

One of the most extensive analyses is that of Corradi \& Schwarz (1995,
hereafter CS95), who considered 400 PNe with high-quality optical images and
studied in detail the correlation of the morphological properties with
several other physical properties, giving particular emphasis to the
comparison between the elliptical ($e$) and bipolar ($b$) PNe.  I stress once
more the quite strict definition of \bpn\ according to the classification of
Schwarz et al. (1993): these are elongated, axially
symmetric PNe distinguished by an ``equatorial'' waist from which two faint,
extended lobes depart. CS95 found that \bpn\ have:\\
{\it a)} a scale height on the Galactic plane of
130~pc compared to the value of 320~pc for the $e$
objects and of 260~pc for the global sample of Galactic disk PNe;\\
{\it b)} smaller deviations than the other morphological types from pure circular 
Galactic rotation;\\
{\it c)} the hottest central stars among PNe;\\
{\it d)} chemical overabundances of helium and nitrogen;\\
{\it e)} outflow velocities up to an order of magnitude greater than the
typical expansion velocities of PNe;\\
{\it f)} giant dimensions;\\
{\it g)} different ``evolutionary''  tracks in the two-color IRAS diagram.

The above properties---especially {\it a,b,c,d}---indicate that {\bpn\ are
produced by more massive progenitors than the other morphological classes}.
These results were later confirmed by other authors (e.g. Gorny, Stasinska, \&
Tylenda 1997; see also Manchado, this volume).

In this contribution, I will add some comments to  the discussion of those
properties which appear to be most important for understanding
 the formation of the
different morphological classes I then discuss whether these
 put real constraints on
the models and present some other recent results on \bpn.

\begin{figure}
\plotone{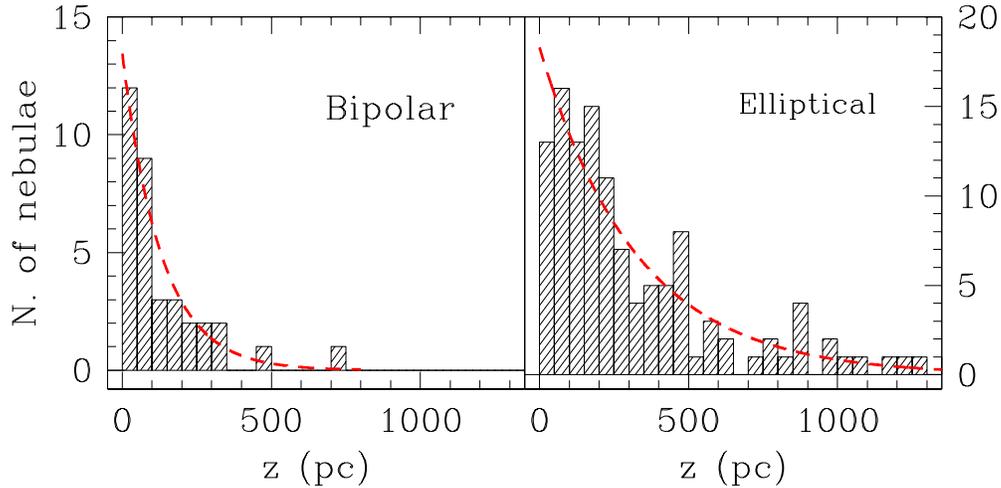}
\caption{The $z$ distribution of $b$ and $e$ PN from CS95. The dashed lines
indicate the best exponential fits with scale heights as indicated in the
text.}
\end{figure}

\section{The mass of  \bpn\ progenitors}

As mentioned above, the conclusion that \bpn\ have more massive progenitors
than \epn\ is consistent with several properties of the nebulae and their
central stars. But the only property which allows us quantitatively to derive 
mass limits for the progenitors of the different morphological classes is the
distribution of the nebulae along the vertical ($z$) direction of the
Galactic disk.  The method is to compare the $z$-distribution of PNe with
that of main-sequence stars in the Galaxy.  To do this, an exponential-in-$z$
distribution function is usually assumed, $n(z)=n_0e^{-z/z_h}$, where $z_h$
is the scale height.  Note that this distribution function has the convenient
property that \zmedio$\equiv\sigma\equiv z_h = 1.44
$$\langle $$|z|$$\rangle_{1/2}$, where \zmedio, $\sigma$, and
$<$$|z|$$>_{1/2}$ are the mean, the standard deviation and the median,
respectively, so that one would be tempted to derive $z_h$ by simply
measuring \zmedio.  The exponential function which a scale height independent
from the galactocentric distance appears to be a good representation for the
distribution of light (star volume density) in spiral galaxies (Wainscoat et
al. 1992).

The $z$-distribution of 35 \bpn\ and 119 \epn\ in CS95 is shown in Figure~1.
The histograms are indeed fairly well fitted by exponential distributions, with
$z_h$ = 130~pc and $z_h$ = 320~pc for the $b$ and $e$ types, respectively. CS95
also computed $z_h$ = 260~pc for the whole sample of Galactic disk PNe. These
figures should be compared to the scale heights of main-sequence stars in the
Galaxy, which as shown in Fig.~2 are basically divided into two regimes:
$z_h$ around 100~pc for spectral types earlier than F5, and around 300~pc for
types later than F8.  The increase of $z_h$ with spectral type is thought to
be an age effect, caused by the so-called ``dynamical heating'' of the disk
(Wielen et al. 1984), which would produce a continuous growth of the
$z$-velocity dispersion of stars.  From Fig.~2, it is clear that \bpn\ are
associated with the low $z_h$ regime, while \epn\ with the high $z_h$
one. This corresponds to limits of $m>1.3$~\sm\ for the initial mass of the
progenitors of \bpn, and $m<1.3$~\sm\ for the \epn. These figures are only
slightly different from those quoted by CS95, being slightly more
conservative according to the discussion below of the uncertainties
involved in their derivation.
\begin{figure}
\plotone{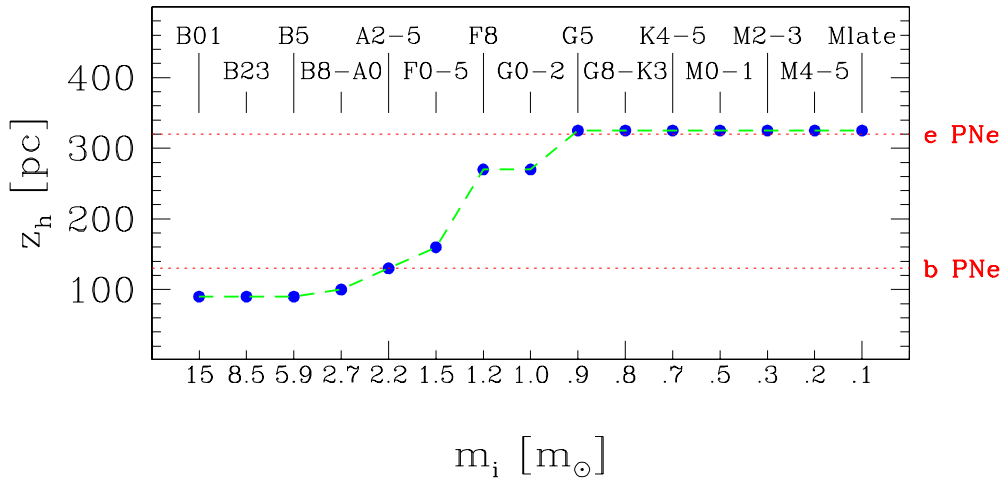}
\caption{The scale height above the Galactic plane for different groups of
main sequence stars in the Galaxy (data from Wainscoat et al. 1992).  The
dotted lines indicate the scale heights of \bpn\ and \epn\ as from the fits
in Figure~1.}
\end{figure}

\subsection{Caveats}
In deriving mass limits for the progenitors of the different morphological
classes from the Galactic $z$ distribution, one should consider the
following points:
\begin{itemize}
\item
The main limitation in this kind of analysis is the uncertainty in the
distances of individual PNe. Considering that \bpn\ are a very peculiar class
of nebulae, it would not be surprising if their distances obtained via the
usual statistical methods were affected by systematic errors. It would be
sufficient that their distances were underestimated by a factor of two to
raise serious doubts about the existence of a real difference in their
Galactic scale height with respect to the whole sample of PNe.  To remove
this doubt, I have estimated {\it kinematical} distances for the sample of
\bpn\ in CS95. As is widely known, the method consists in assuming that the
objects participate in the general circular rotation around the Galactic
center. Their distances are then estimated by comparing their apparent radial
velocities with those expected for circular orbits as a function of 
heliocentric distance and Galactic longitude. Errors in the distances
provided by this method are large, but it remains the only method
 which, at present,
can be applied to many \bpn, and which is free from risk of systematic
errors.  A comparison of the kinematical distances with those adopted by CS95
for 30~\bpn\ is plotted in Figure~3. Apart from an obvious large
dispersion of points, from this plot we can exclude the possibility
that the distances of
\bpn\ were underestimated by CS95 by a factor of two or more, since in that
case they would preferentially lie close to the 2:1 relation (dashed line),
while most of them are instead found close to the 1:1 relation (solid line).
In fact, computing the scale height of the 30~\bpn\ using the kinematical
distances, yields $z_h\sim$150~pc.  Thus this exercise fully confirms the
results of CS95.
\begin{figure}
\plotone{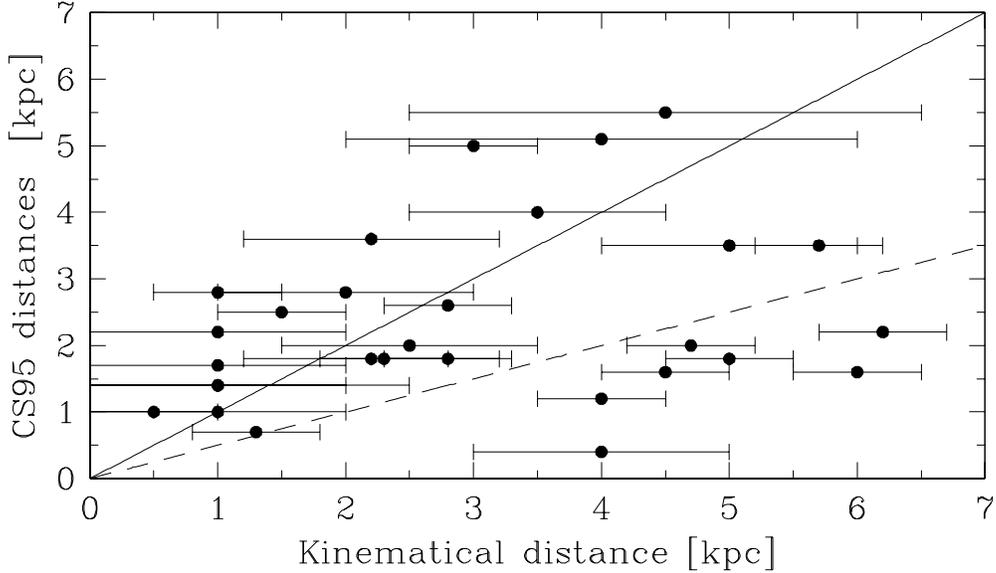}
\caption{Kinematical distances vs. CS95 distances  for
30~\bpn. Estimated errors for the kinematical distances are indicated, as
well as the 1:1 (solid line) and 2:1 (dashed one) relations.}
\end{figure}
\item
When determining the scale height of a class of objects, one has to be careful
with the sample size. The number of objects should be large enough to check
whether the exponential model distribution is a fair representation
of the observed one, and that it is sufficiently well sampled.   
The derivation of $z_h$ directly from \zmedio\ for small samples should also
be avoided.
\item
As discussed above, main-sequence stars in the Galaxy are basically divided
into ``early-type'' stars with $z_h\sim$100~pc and ``late-type'' objects with
$z_h\sim$300~pc, the rapid transition to larger scale heights occurring
between spectral classes F0 and G5.  Considering the uncertainties involved
in the derivation of the scale height for PNe, and in particular the present
poor
knowledge of distances, it is therefore not possible to derive precise mass
values for their progenitors. Note also that the reference samples of stars
are a mixture of main-sequence objects of different ages (recently born or on
the point of leaving the main sequence) while the PNe are clearly all very
evolved objects. A population of PNe is then expected to lie at higher $z$
than a (younger) random mixture of its main-sequence progenitors, provided
that the main-sequence lifetime is sufficiently long to allow the dynamical
heating of the disk to act. This introduces an additional uncertainty into the
analysis.  For these reasons, it is not advisable to derive more accurate
masses for PN progenitors than those which result from dividing them in the
above two scale height regimes.  Taking a mass value in the middle of the
transition from low to high scale heights for main-sequence stars, say
1.3~\sm\ corresponding to spectral types F5-F8, the only robust conclusion is
therefore that the progenitors of \bpn\ have initial masses greater than this
value, while those of \epn\ have smaller ones.
\end{itemize}

\subsection{A real constraint?}

Although the difference in progenitor masses of $b$ and $e$ PNe is a very
important constraint that must be taken into account when explaining their
formation, at present it does not appear to be able to discriminate as to
 whether
these morphological classes are produced by single or binary systems, which
is one of the main topics of this conference.

The fact that \bpn\ have  higher-mass progenitors is the basic starting
point for {\it single} star theories.  Garc\'\i a-Segura et al. (1999)
conjecture that, if the initial mass is larger than $1.3$~\sm\ (note that
this limit coincides with that derived in the previous section for \bpn),
stars might rotate on the AGB with a velocity close to the critical one,
causing a markedly aspherical mass deposition. This is because their cores would
not have been spun down onto the main sequence (as occurs for lower-mass stars),
evolve decoupled from the envelope, and remain fast rotators until angular
momentum is redistributed on the AGB, spinning up the outer layers.  Rotation
would force the envelope ejection in the equatorial plane of the AGB star,
providing the correct mass-loss geometry needed to form a bipolar PN.

But  {\it binary} models also have an explanation for the higher progenitor
masses of \bpn. In fact, according to Soker (1998) a low-mass star in a
binary system would interact strongly with the companion already on the RGB
(the ratio of the stellar radii during the AGB and the RGB,
$R_{\rm AGB}$/$R_{\rm RGB}$, is small for low-mass stars), causing an enhanced mass-loss which would leave a low-mass and small AGB envelope. Then the
interaction with the companion on the AGB would be weak preventing the
formation of a $b$ PN. On the other hand, higher-mass stars, for which
$R_{\rm AGB}$$\gg$$R_{\rm RGB}$, would mainly interact on the AGB, and because of
this interaction would be able to produce a $b$~PN.

In addition, it should be mentioned that Mellema (1997) showed that in
low-mass stars the post-AGB evolution is slow enough for the ionization
front to have time to smooth out the density contrast between the equatorial
plane and the polar directions, preventing the formation of a bipolar PN even
if the AGB mass-loss geometry were the favorable one. In higher-mass stars,
the post-AGB evolution is so fast that the density distribution is 
unchanged. It should be calculated, however, whether this conclusion still
holds for clumpy distributions of gas.

\section{Chemical abundances of bipolar PNe}

The association of \bpn\ with the chemical type I, i.e. with He- and/or N-rich
objects, has been known since Peimbert's 1978 paper. CS95 discussed the amount of data
available in the literature in 1995, and more recently their results were
refined by means of new observations.  Corradi et al. (1997b) and Perinotto \&
Corradi (1998) made a detailed spatially resolved study of the abundances of
a sample of 15 \bpn with the following main results:
\begin{itemize}
\item
within the errors, the nebulae are {\it chemically homogeneous} in He, O, N;
\item
Ne, Ar, and S abundances have systematic increases toward the outer regions of
the nebulae. This might be due to errors inherent to the use of the standard
{\it icf} method (Alexander \& Balick 1997);
\item
it is confirmed that \bpn\ are overabundant in He and N (type I) with respect
to elliptical PNe, as indicated in Table~1. This is ascribed to efficient
second and third dredge-up and burning at the base of the convective envelope
in the most massive progenitors;
\begin{table}
\caption{Average abundances (by number). 
The number of objects is given in parentheses.}
\begin{center}
\begin{tabular}{lllll}
\hline\hline\\[-10pt]
 & He/H & O/H$\times10^{-4}$ & N/O & N/H$\times10^{-4}$\\
\hline\\[-10pt]
\bpn\      & 0.15 (27)& 3.97 (27)& 1.24 (28)& 4.17 (27)\\
\epn\      & 0.11 (66)& 5.03 (68)& 0.40 (61)& 2.10 (60)\\ 
Sun        & 0.10     & 8.51     & 0.12   & 1.02    \\
\hline
\end{tabular}
\end{center}
\end{table}
\item
the highest He overabundances displayed by some \bpn\ (e.g. M~3-2, He~2-111,
and NGC~6537) cannot by reproduced by any current model of AGB evolution;
\item
oxygen depletion is suggested for the nebulae with the highest N/O
abundances, indicating that an efficient ON cycle process has occurred in their
progenitors.
\end{itemize}

\subsection{Any problem with the binary models?}

Let's now turn our attention to ``massive'' stars. Soker (1998) has argued
that 30--40\% of massive stars (M$_i\ge2.3$~\sm) have companions in the right
separation range to form bipolar ($b$) PNe. The other 60--70\% would form
elliptical ($e$) nebulae.  If evolution in a binary system do not affect the
chemical enrichment of the stars, then one expects to find, at any abundance
interval in the range for massive PNe progenitors, twice as many $e$ as $b$
PNe (assuming the same nebular life time).  In our sample, after correcting
for sample size effects, for large He and N abundances we have instead the
reverse situation, with more $b$ than $e$ PNe.
This raises the question of where all the non-bipolar PNe
with large He and N abundances are (the expected progeny of single massive
stars)?

An explanation within the binary scenario would require {\it
extra enrichment} caused by the binary interaction.  One hint in this sense
is the fact that the highest He abundances of \bpn\ are not reproduced by any
current model for single stars.

\section{Orientation in the Galaxy of axisymmetrical PNe}

There were some strong suggestions in the past that the symmetry axis of
aspherical PNe were not oriented randomly within the Galaxy. In particular,
Melnick \& Harwit (1975) and recently Phillips (1997) have claimed that axially
symmetrical PNe have their axes preferentially inclined at low angles to
the Galactic plane.  Such a property would put very interesting constraints
on the mechanisms producing the asphericity in PNe. In other classes of
objects, as in SN remnants (Gaensler 1998) and nebulae ejected by massive
stars (Hutsemekers 1999), the nebular orientation, for instance, was  found to
be clearly related to the Galactic magnetic field.

Based on a larger and more homogeneous sample, and on a more rigorous
statistical analysis than in the previous studies, Corradi et al.
(1997a) showed instead that there is no strong evidence for such an
alignment, at least partially  removing interest to the issue.

\section{The role of detached binary systems: symbiotic stars}

Finally, I conclude with a brief comment on the observations of binary
stars in PNe and their relation with \bpn. A review of very close
binaries, with periods of a few hours to a few days, is presented by H. Bond
in this volume. These objects are expected to have undergone a common envelope
phase during the AGB evolution of the star which has now ejected its PN.
Although most PNe with a close binary nucleus are aspherical, only a
relatively small fraction of them (some 25\%) has a truly bipolar nebula.

In my opinion, the situation looks more promising for wider interacting
binaries which avoid the common envelope phase, as also supported by Soker
(1997).  In this respect, known detached binaries like symbiotic stars give a
practical demonstration of the ability of these systems to form bipolar
nebulae, rings and jets. In symbiotic stars containing a Mira, all nebulae
are markedly aspherical, and about a half are bipolar/ring.  This suggests
that the kind of interactions occurring in these systems (formation of
collimating accretion/excretion disks, production of collimated fast winds
from the hot components) might provide the explanation of the collimation of
the outflows in \bpn. A thorough discussion of the nebulae around symbiotic
stars and of their similarities with \bpn\ can be found in the contributions
of Corradi et al. and Mikolajewska (this volume), as well as in Corradi et
al. (1999a, 1999b).


\end{document}